\documentclass{birkjour}

\usepackage{amssymb}
\usepackage{cite}
\allowdisplaybreaks
\begin{document}

\title[The reasonable effectiveness of deformation theory in physics] 
{The reasonable effectiveness\\
of mathematical deformation theory\\ in physics}    
\thanks{\copyright{\footnotesize {2017 Daniel Sternheimer}}}

\author{Daniel Sternheimer}
\address{Department of Mathematics\\
Rikkyo University\\
Tokyo, Japan\\
and\\
Institut de Math\'ematiques de Bourgogne\\
Dijon, France} 



\begin{abstract}
This is a brief reminder, with extensions, from a different angle and for a 
less specialized audience, of my presentation at WGMP32 in July 2013, to which 
I refer for more details on the topics hinted at in the title, mainly
deformation theory applied to quantization and symmetries (of elementary 
particles).  
\end{abstract}
\maketitle

\section{Presentation}
In 1960 Wigner \cite{Wi60} marveled about ``the unreasonable 
effectiveness of mathematics in the natural sciences." In 1963/64 appeared 
index theorems for pseudodifferential operators. I participated (expos\'e 23,
\cite{CS64}) in the Cartan--Schwartz seminar that developed the announcement of 
the first result (a parallel seminar was held in Princeton under Richard Palais, 
and different proofs and extensions were published by Atiyah et al. a few years later). 
In 1964 appeared Gerstenhaber's theory of deformations of algebras \cite{Ge64}. 
[Murray Gerstenhaber is now 90 and still active.]
Soon after Gerstenhaber's seminal work it was realized that, from the viewpoint 
of symmetries, special relativity is a deformation.

The underlying idea is that new fundamental physical theories can, so far 
a posteriori, be seen as emerging from existing ones via some kind of 
deformation. The main paradigms are the physics revolutions from the beginning 
of the twentieth century, special relativity (symmetry deformation from the 
Galilean to the Poincar\'e groups) and quantum mechanics (via deformation 
quantization).

Indeed in the mid-seventies all this converged to explain quantum mechanics 
as a deformation of classical mechanics \cite{BFFLS}, what is now called 
deformation quantization. Quantum groups and noncommutative geometry can be 
considered as avatars of that framework. In this very short paper 
(see \cite{dsWGMP32}, and references therein, for more details on most of 
these ideas) I present these notions in general terms, describe some results. 
and indicate some perspectives, including suggestions to use the framework to 
put on ``non-clay feet" the ``colossal"  standard model of elementary particles, 
and maybe explain the ``dark universe."

\subsection{The problem}
``It isn't that they can't see the solution. It is that they can't see the 
problem." That is a quote from a detective story by G. K. Chesterton (1874--1936) 
(``The Point of a Pin" in ``The Scandal of Father Brown" (1935)).
The problem, in my view, is that \textsl{the Standard Model of elementary 
particles could be a colossus with clay feet}. Cf. in the Bible (Daniel 2:41-43), 
the interpretation by Belteshazzar (a.k.a. prophet Daniel) of Nebuchadnezzar's 
dream.

The physical consequences of the approach described here might be 
revolutionary but in any case there are, in the mathematical tools required to 
jump start the process, potentially important developments to be made.

\subsection {Some motivating quotes}
\textsl{Albert Einstein}: ``The important thing is not to stop questioning. 
Curiosity has its own reason for existing."\\
``You can never solve a [{\small fundamental, precision added by DS}] problem 
on the level on which it was created."\\
\textsl{Gerard 't Hooft} (about Abdus Salam) \cite{tH08}: ``To obtain the 
Grand Picture of the physical world we inhabit [\ldots] courage is required. 
Every now and then, one has to take a step backwards, one has to ask silly 
questions, one must question established wisdom, one must play with ideas like 
being a child. And one must not be afraid of making dumb mistakes. By his 
adversaries, Abdus Salam was accused of all these things."\\
\textsl{Eugene Paul Wigner} \cite{Wi60}: ``Mathematical concepts turn up in 
entirely unexpected connections. Moreover, they often permit an unexpectedly 
close and accurate description of the phenomena in these connections.
Secondly, just because of this circumstance, and because we do not understand 
the reasons of their usefulness, we cannot know whether a theory formulated 
in terms of mathematical concepts is uniquely appropriate."\\
\textsl{Sir Michael Atiyah} (at ICMP London 2000, \cite{Ati00}): 
``Mathematics and physics are two communities separated by a common language."\\
\textsl{Paul Adrien Maurice Dirac} \cite{Di49}: ``Two points of view may be 
mathematically equivalent [\dots] But it may be that one point of view may 
suggest a future development which another point does not suggest [\ldots] 
Therefore, I think that we cannot afford to neglect any possible point of 
view for looking at Quantum Mechanics and in particular its relation to 
Classical Mechanics."

\section{Headlines}
A scientist should ask himself three questions: Why, What and How. Of course, 
work is 99\% perspiration and 1\% inspiration. Finding how is 99\% of the 
research work, but it is important to know what one is doing and even more 
why one does such a research.

What we call ``physical mathematics" can be defined as mathematics inspired 
by physics. While in mathematical physics one studies physical problems with 
mathematical tools and (hopefully) rigor. [Theoretical physics uses 
mathematical language without caring much about rigor.] In addition, as to 
``what" and ``how" to research there are important differences between 
mathematicians and physicists. Indeed, even when taking their inspiration from 
physics (which fortunately is again often the case now, see e.g. \cite{Ati00}), 
mathematicians tend to study problems in as general a context as possible, 
which may be very hard. But when the aim is to tackle specific physical 
problems, and though generalizations may turn out to have unexpected
consequences, it is often enough to develop tools adapted to the desired 
applications. 

That is the spirit which inspired the approaches Moshe Flato and I developed 
(with coworkers of course) since the mid 60s, and which I am continuing this 
millenium. In what follows I give a flavor of two main topics we tackled, i.e.
original applications both of symmetries to particle physics and of 
quantization as a deformation, and of the combining program that I am now 
trying to push forward for the coming generation(s). 

\subsection{Deformation quantization and avatars}
As we said above the two major physical theories of the first half of the
twentieth century, relativity and quantization, can now be understood as based 
on deformations of some algebras. That is the starting point of Moshe Flato's
``deformation philosophy". Deformations (in the sense of Gerstenhaber 
\cite{Ge64}) are classified by cohomologies. The deformation aspect of 
relativity became obvious in 1964, as soon as deformation theory of algebras 
(and groups) appeared, since one can deform the Galilean group symmetry of 
Newtonian mechanics $SO(3)\rtimes\mathbb{R}^3\rtimes\mathbb{R}^4$ to the 
Poincar\'e group $SO(3,1)\rtimes\mathbb{R}^4$. 

Though (when Moshe arrived in Paris) I studied in \cite{CS64} the composition
of symbols of elliptic operators, and in spite of the fact that the idea that 
some passage from classical to quantum mechanics had been ``in the back of the 
mind" of many, it took a dozen more years before quantization also could be 
mathematically understood as a deformation, with what is now called deformation 
quantization, often without quoting our founding 1978 papers \cite{BFFLS}. 
(These have nevertheless been cited over 1000 times if one includes the physics 
literature, and so far 271 times for paper II in the mathematics literature, 
according to MathSciNet.) 

Explaining the process in some detail would be beyond the scope of this short 
overview, so we refer to \cite{dsWGMP32} and references therein. In a nutshell 
the idea is that, instead of a complete change in the nature of observables 
in classical mechanics (from functions on phase space, a symplectic or Poisson 
manifold, to operators on some Hilbert space in quantum mechanics), the algebra 
of quantum mechanics observables can be built on the same classical observables 
but with a deformed composition law (what we called a ``star product" since the
deformed composition law was denoted~$\star$). The main paradigm is the 
harmonic oscillator. That can be extended to field theory (infinitely many 
degrees of freedom), and more.  

Related representations of Lie groups can then be performed on functions on 
orbits with deformed products, instead of operators. Similarly, though they 
arose in quite different contexts in the 80s, based on previous works from 
the 70s by their initiators (Faddeev's Leningrad school for quantum groups 
coming from quantum integrable models, and Alain Connes' seminal works on
von Neumann algebras for noncommutative geometry), both can be (see e.g. 
\cite{Dr86,CFS92}) considered as avatars of our framework.    

In particular, in the ``generic case", quantum groups are deformations of
an algebra of functions on a Poisson Lie group or of a dual algebra 
\mbox{\cite{BFGP, BGGS}}. The idea was extended in the 90s to multiparameter 
deformations (with commuting parameters), and (unrelatedly) to the case when 
the parameter is a root of unity -- in which case the deformed Hopf algebra is 
finite dimensional, with finitely many irreducible representations. 
That idea has not yet been extended to multiparameter deformations at roots 
of unity, nor a fortiori to noncommutative parameters (a notion which
is not part of Gerstenhaber's approach and is not yet defined).  
 
\subsection{Symmetries of elementary particles}
A posteriori one can say that the geometric aspect of deformation theory was 
known in physics since the antiquity, in particular when (in the fifth 
century B.C.) Pythagoras conjectured that, like other celestial bodies, the 
earth is not flat; two centuries later Aristotle gave phenomenogical indications 
why this is true, and ca. 240 B.C. Eratosthenes came with an experimental 
proof of the phenomenon, giving a remarkably precise evaluation of the radius
of the ``spherical" earth. In mathematics one had to wait for Riemann's surface
theory to get an analogue.  
 
In another context, in the latter part of last century arose the standard 
model of elementary particles, based on empirically guessed symmetries. The 
untold rationale was that symmetries are important to explain the spectra 
observed in atomic and crystalline spectroscopy (as shown e.g. in Moshe Flato's 
M.Sc. thesis under Racah \cite{KibCMF}). There one knows the forces and 
symmetries make calculations feasible. In nuclear spectroscopy, the subject of 
Moshe's PhD thesis under Racah (which he never defended because Racah died 
unexpectedly in Firenze on his way to meet Moshe in Paris, and by then Moshe 
had already completed a D.Sc. in Paris), symmetries can be used as ``spectrum
generating". 

That is also how symmetries were introduced empirically in particle physics, 
starting with isospin $SU(2)$ since the 30s, then with a rank 2 compact Lie 
group (thoroughly studied in the less known \cite{BDFL62}) after ``strangeness", 
a new quantum number, was introduced in the 50s (in particular thanks to people 
like Murray Gell-Mann, who was then daring to tackle unpopular topics, before 
becoming a kind of ``guru" for at least a generation). That was quickly 
restricted to ``flavor" $SU(3)$. 

A natural question was then what (if any) is the relation between such 
``internal" symmetries and the ``external" symmetries like the Poincar\'e group, 
in particular to explain the mass spectrum inside a multiplet. In 1965 we 
objected \cite{FS65} to a ``no-go theorem" claiming to show that the connection 
must be a direct sum, giving counterexamples shortly afterward. One should be 
careful with no-go theorems in physics, which often rely on unstated hypotheses. 
[Another issue, which was not so explicitly mentioned, is e.g. how one can 
have the 3 octets of the ``eightfold way", based on the same adjoint 
representation of $SU(3)$, associated with bosons of spin 0 and 1, and 
fermions of spin $\frac{1}{2}$.]

But what (if anything) to do with the basic 3-dimensional representations of 
$SU(3)$? In 1964 Gell-Mann (and others) came with the suggestion to associate 
them with ``quarks", hypothetical entities of fractional charge which (being 
``confined" and of spin $\frac{1}{2}$) cannot be directly observed nor coexist 
e.g. in a hadron (strongly interacting particle). Initially there were three 
``flavors" of quarks. Later consequences of the quark hypothesis were observed 
and we now have 3 generations (6 flavors) of quarks. To make possible their 
coexistence in a hadron they were given (three) different colors, whence 
``color $SU(3)$." Soon, on the basis of that empirically guessed symmetry, 
in a process reverse to what was done in spectroscopy, dynamics were developed, 
QCD (quantum chromodynamics) with a non-abelian gauge $SU(3)$ on the pattern 
of QED (quantum electrodynamics, with abelian $U(1)$ gauge). 
The rest is history but not the end of the story.

Already in 1988 \cite{FF88}, Flato and Fronsdal explained how, if one
deforms space-time from Minkowski to Anti de Sitter (also a 4-dimensional
space-time but with tiny negative curvature) and as a consequence the 
Poincar\'e group to AdS $SO(2,3)$, one can explain the photon as dynamically 
composite of two Dirac ``singletons" (massless particles in 1+2 dimensional 
space-time) in a way compatible with QED. That was an instance of the AdS/CFT 
correspondence which we had detailed e.g. in 1981 \cite{AFFS}. 

After numerous of papers on singleton physics, in Moshe's last paper 
\cite{FFS99} we described how the AdS deformation of Poincar\'e may explain 
the newly discovered neutrino oscillations, which showed that neutrinos are 
not massless. Going one step further, shortly afterward Fronsdal \cite{Fr00} 
explained how, on the pattern of the electroweak model and on the basis 
of AdS deformation, the leptons (electron, muon, tau, their antiparticles 
and neutrinos) can be considered as composites of singletons, initially 
massless, massified by 5 Higgs. (This predicts e.g. 2 new ``W and Z like" 
bosons.)

\subsection{Combining both, and perspectives}
In line with our deformation philosophy, the idea is that the question of 
connection between symmetries could be a false problem: the ``internal" 
symmetries on which the Standard Model is based might ``emerge" from the 
symmetry of relativity, first by ``geometric" deformation (to Anti de Sitter, 
with singleton physics for photons and leptons) followed (for hadrons) by a 
quantum group deformation quantization.

The program to deform Poincar\'e to AdS for particle symmetries was developed 
by Flato and coworkers since the 70's (see e.g. \cite{AFFS,FF88,FFS99} and 
references therein), in parallel with deformation quantization. 
In the beginning of this millenium it dawned on me that the two could 
(maybe should) be combined by deforming further AdS. The natural way to do
so is within the framework of quantum groups, possibly multiparameter 
since we now have~3 ``generations" of elementary particles, which as a fringe 
benefit might make room for the traditional (compact) internal symmetries.
In view of their special properties, quantum groups at roots of unity seem a 
promising structure, Of course such an approach is at this stage merely a
general framework to be developed, and would be only part of the picture if 
one cannot ``plug in" the present QCD dynamics.     

In particular with for parameters the algebra of the Abelian group 
$\frac{\mathbb{Z}}{3\mathbb{Z}}$, at e.g. sixth root of unity, one might 
be able to recover $SU(3)$ and ``put on solid ground" the Standard Model. 
Or maybe replace its symmetry with a better one, requiring to ``go back to
the drawing board" and re-examine half a century of particle physics (from the
theoretical, phenomenological and experimental viewpoints). The former 
alternative is relatively more economical. The latter requires huge efforts,
albeit without having to build new machines, which Society is unlikely to 
give us.  

That raises hard mathematical problems. (E.g. the tensor product of two
irreducible representations of a quantum group at root of unity is 
indecomposable.) A solution to part of these has an independent mathematical 
interest. Combined with the necessary detailed phenomenological study required, 
that might lead to a re-foundation of half a century of particle physics. 

There could also be implications in cosmology, including a possible 
explanation of dark matter and/or of primordial black holes, which were 
introduced already in 1974 by Hawking \cite{CH74} and are now considered as 
a possible candidate for ``dark matter" (see e.g. \cite{Ca17}). Since
that is still an open question (in contradistinction with elementary particles 
for which there is a solution, even if the problem is occulted) the community 
might be more receptive to try such ideas.     

\subsection*{Acknowledgment}
This short panorama benefited from 35 years of active collaboration with 
Moshe Flato (and coworkers), from close to \verb+\infty+ discussions with 
many scientists (too numerous to be listed individually) and from the 
hospitality of Japanese colleagues during my so far over 14 years of relatively 
generous retirement from France, after 42 years in CNRS.

\end{document}